\documentclass[baaa]{baaa}

\usepackage[pdftex]{hyperref}
\usepackage{subfigure}
\usepackage{natbib}
\usepackage{helvet,soul}
\usepackage[font=small]{caption}

\contriblanguage{1}

\contribtype{1}

\thematicarea{4}

\title  {Embedded clusters: upgrading visual and infrared photometric 
analysis with Gaia DR2 and ASteCA}

\titlerunning{Embeded clusters}

\author{E.E. Giorgi\inst{1,2}, G.R. Solivella\inst{1,2}, A. Cruzado\inst{1,2}, R.A. Vázquez\inst{1,2}, G.I. Perren\inst{2} 
\& T. Canavesi\inst{3}}
\authorrunning{Giorgi et al.}

\contact{egiorgi@fcaglp.unlp.edu.ar}

\institute{ Facultad de Ciencias Astron\'omicas y Geof\'isicas de La Plata, UNLP, Argentina \and 
Instituto de Astrof\'isica de La Plata, IALP (UNLP-CONICET), Argentina \and Instituto de Física 
de La Plata, IFLP (UNLP-CONICET)}

\resumen{Los cúmulos embebidos son grupos de estrellas que no se han separado aún de la nube 
original donde se han formado, por lo que es fundamental obtener las distancias precisas y las propiedades 
de estos grupos. Se presentan los resultados de cinco cúmulos embebidos: [DBS2003]5, [DBS2003]60, [DBS2003]98, 
[DBS2003]116 y [DBS2003]117. Los resultados provienen de una combinación de fotometría CCD UBVI profunda, adecuada 
para identificar estrellas azules débiles, e información de relevamientos infrarrojos disponibles. A su vez la fotometría 
fue relacionada con los datos de movimientos propios y paralajes provenientes de Gaia DR2. Cada objeto fue tratado en 
un espacio múltiple usando el código ASteCA, especialmente diseñado para realizar un análisis automático y proveer los 
parámetros fundamentales de los grupos de estrellas en caso de que se trate de un cúmulo.

}

\abstract{Embedded clusters are groups of stars which have not dispersed yet the residual of the parental cloud 
where they were born so getting precise distances and properties of
these groups turns out to be an essential task. We present results for five embedded clusters: [DBS2003]5, 
[DBS2003]60, [DBS2003]98, [DBS2003]116 and [DBS2003]117. Results come from a combination of deep CCD UBVI photometry 
suitable to identify blue faint stars and infrared information from available surveys. In turn photometry was linked 
with proper motions and parallaxes from Gaia DR2. Each object was treated in a multi-space using ASteCA, a code especially 
designed to perform an automatic data analysis and aimed at providing the fundamental parameters of star groups, in case 
they compose a real entity. 

}

\keywords{(Galaxy:) open cluster and associations: individual (Havlen-Moffat \#1)--- stars massive --- stars Wolf-Rayet --- stars early type
}
\begin{document}
\maketitle

\section{Introduction}
\label{S_intro}
Embedded clusters are groups of stars which have not dispersed yet the residual of the parental cloud where they 
were born. Therefore, they are fully or partially obscured due to their insertion in the parental cloud where the 
process of star formation has recently taken place. Accordingly, an embedded cluster may host very young stars still 
gravitationally dominated by the presence of the molecular cloud. By themselves they are an essential category of 
astronomical objects whose main properties started being unveiled during the last two decades thanks to the advance 
in infrared detectors and new telescope design. However, molecular clouds may geometrically interpose between observers and 
far star groups that appear highly obscured for this reason and not because they are physically related to the molecular cloud.
Therefore, getting precise distances and properties of these groups, mostly faint, highly obscured and in occasions associated 
to small size HII regions, are an essential but challenging task. Fortunately, the Gaia Second data release (Gaia DR2) provides 
us with a huge amount of information, parallax, proper motions and G-photometry, with unprecedented accuracy.  By means of ASteCA, 
a code designed to perform an automatic data analysis, we combine this valuable information with photometric, visual and infrared 
data from other sources. We present results for five candidates to embedded clusters taken from \cite{2003AA...400..533D} 
and \cite{2003AA...404..223B}: [DBS2003]5, [DBS2003]60, [DBS2003]98, [DBS2003]116 and [DBS2003]117. 

\section{Data and Analysis}
Data to be analysed in a multi-space come from Gaia DR2 including parallaxes, photometry (G mag and BP~-RP color index) and proper 
motions, from the 2MASS survey (JHK mag), and also from the UBVI photometry obtained with 2.15m telescope at Complejo Astronómico 
El Leoncito (CASLEO), San Juan. For saving space we only show the analysis performed using Gaia data down to G$\approx$21 mag, 
greatly deeper than the 2MASS limiting magnitude. We applied restrictions to photometric data (G~error $<$ 0.01 mag, 
(BP~-~RP)~error~$<$~0.2 mag) to remove from the analysis stars with the largest uncertainties. No cuts have been applied neither 
on parallax data nor on proper motions data.

\begin{figure*}[htb]
  \centering
  \includegraphics[width=0.99\textwidth]{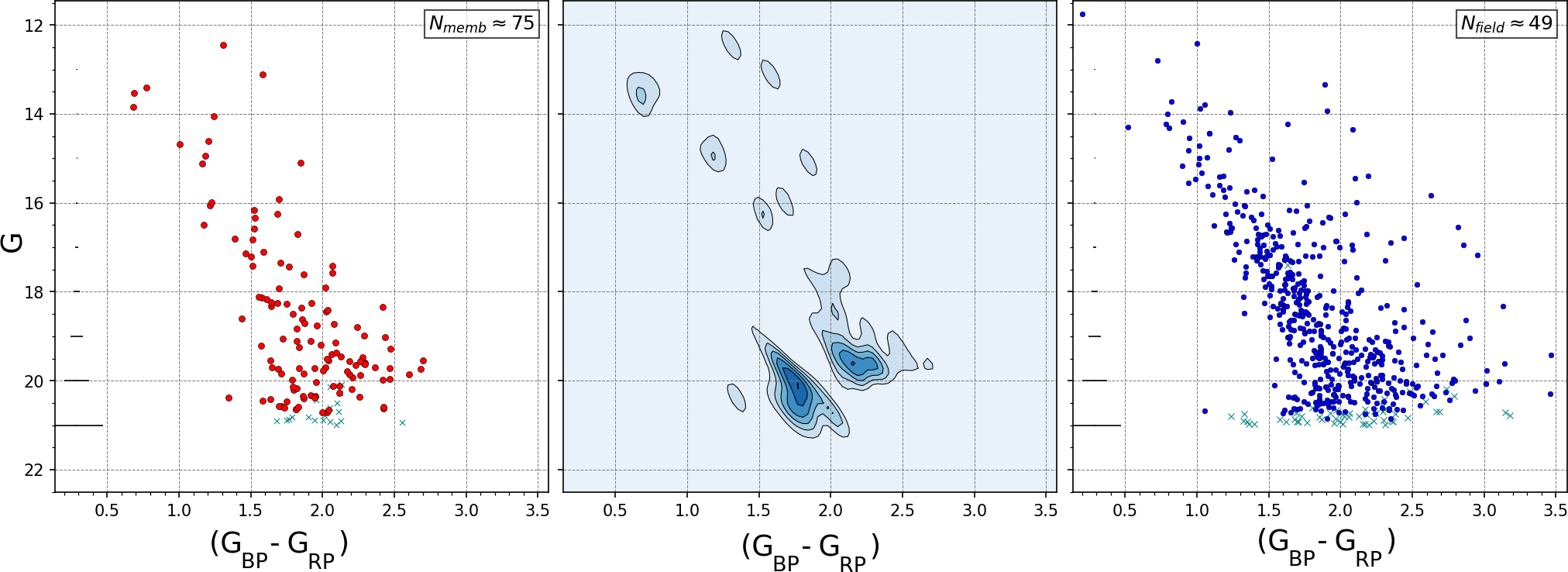}
  \caption{CMDs for the cluster region (left) and the field region (right), with a cleaned cluster KDE in the middle.
}
  \label{Figure1}
\end{figure*}

An analysis of the structural data was performed on the five candidate clusters. The kernel density estimate (KDE) maps is 
obtained using the cluster’s coordinates and a bandwidth value. This process is done to check if the clusters can actually be 
observed as an enhancement over the surrounding field density. Depending on their values, different bandwidths help expose 
relevant features of the observed frame. A clear overdensity is seen in the middle of the frame in [DBS2003]5. In other 
cases (e.g. [DBS2003]116 and [DBS2003]117), this evidence is not so obvious. We warn the reader that an overdensity is not 
necessarily a star cluster.

Below we show in detail the analysis performed for [DBS2003]5. However, Tables \ref{table1} and \ref{table2} summarize the 
results for all groups. Fig. \ref{Figure1} display a comparison of the color-magnitude diagram (CMD) for the cluster region 
(left) and the surrounding field (right). The middle plot is a KDE map showing the result of subtracting the KDE of the field 
region from the KDE of the cluster region. The idea is that this procedure should expose the structure of the actual cluster’s 
sequence (if there is any) in the CMD, by removing the contamination from field stars. No clear cluster sequence can be 
distinguished in the cleaned KDE map, shown in the middle plot of Fig. \ref{Figure1}, even though the number of estimated 
members in this case (obtained by subtracting the field density from the cluster region density) is approximately 75 stars.

\begin{figure}[h!]
  \centering
  \includegraphics[width=0.5\textwidth]{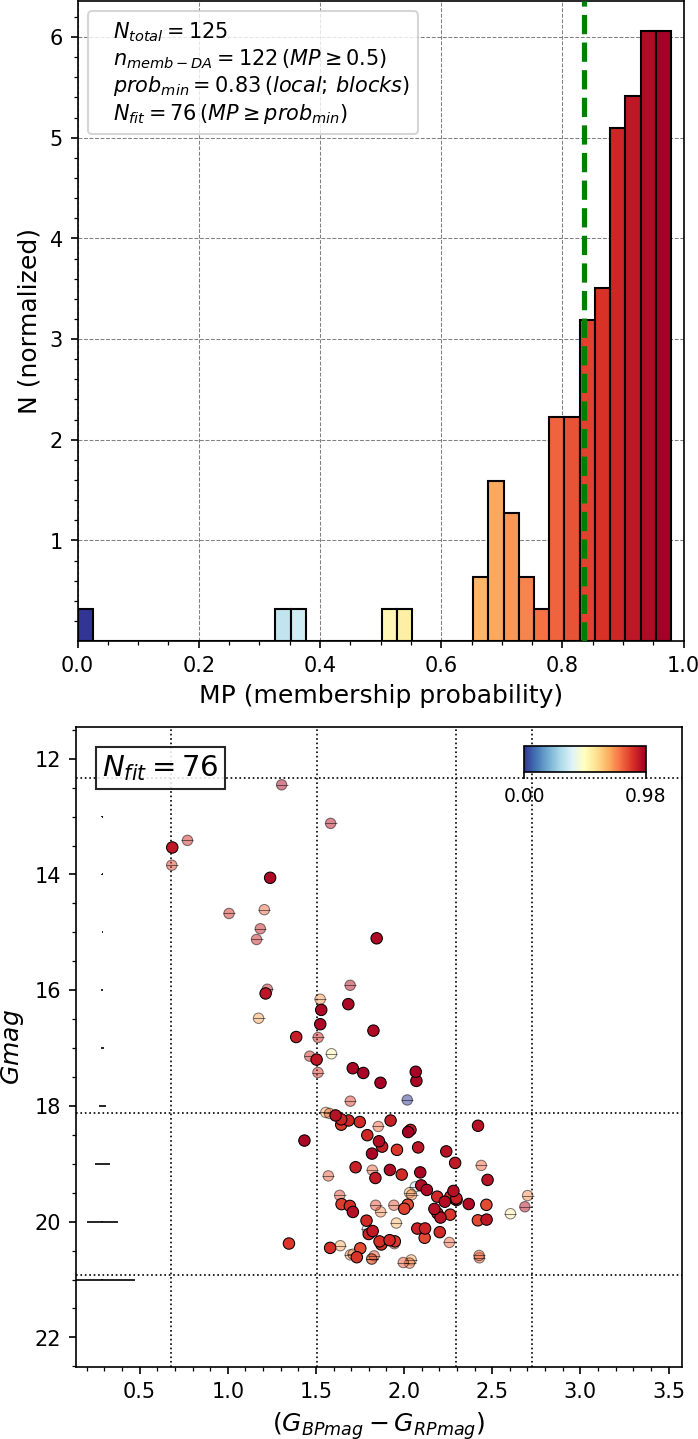}
  \caption{Membership probabilities estimation process.}
  \label{Figure2}
\end{figure}

Following this we applied a membership probability (MP) assigning algorithm to the assumed cluster region (details are 
explained in \cite{2015AA...576A...6P}). The method compares selected features of the stars within the cluster region 
with those in the surrounding field region. In this case, since the Gaia DR2 (BP-RP) color contains large uncertainties, 
parallax and proper motions were employed to estimate the MPs (membership probabilities) for all stars within the assumed 
cluster region. The results of this process are shown in Fig. \ref{Figure2}.

The top panel in Fig. \ref{Figure2} shows the distribution of probabilities assigned to all stars within the cluster 
region. As seen, most values are large and concentrated above MP=0.8. This usually indicates that no clear separation could 
be performed for true cluster members and field interlopers in the respective parallax and proper motion spaces. The middle 
plot in Fig. \ref{Figure2} is a positional chart of those stars within the cluster region colored according to these MPs 
(stars outside the estimated radius are not assigned any probability and hence are drawn as empty circles). Finally the plot 
to the right shows the cluster region’s CMD colored according to the MPs, and with enough stars removed to leave roughly 
the number of members estimated previously. Out of the total 125 stars in the region, 76 are thus kept as the most probable 
members in [DBS2003]5. Again, we see no clear cluster sequence in this cleaned CMD.
\begin{figure}[h!]
  \centering
  \includegraphics[width=0.5\textwidth]{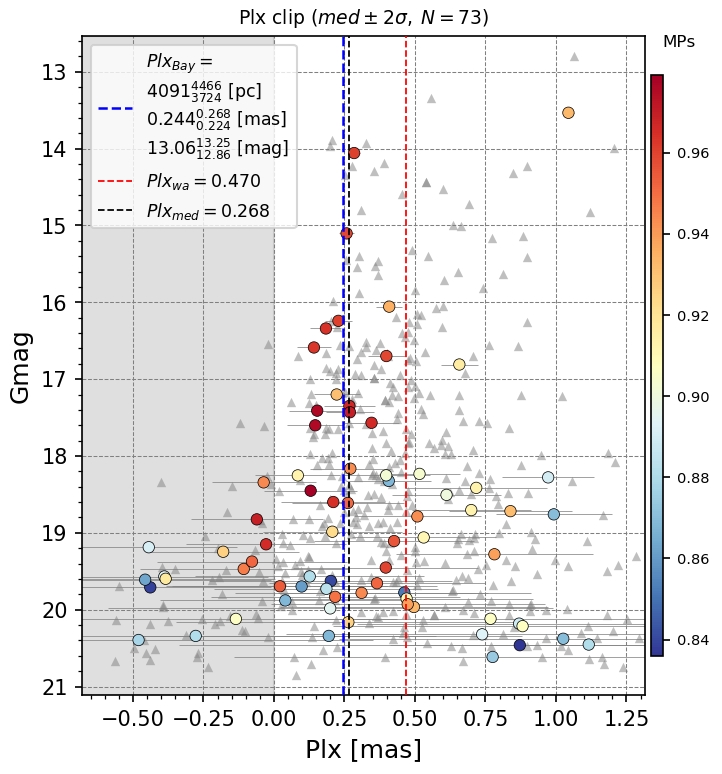}
  \caption{Bayesian analysis performed on the parallax values.}
  \label{Figure3}
\end{figure}

Finally, parallax and proper motions are processed and analyzed to estimate the cluster’s distance and mean proper motion. 
Parallax values were shifted by a +0.029 mas offset, as suggested in \cite{2018AA...616A...2L} and processed with the Bayesian method described
by Bailer-Jones\footnote{
\url{https://github.com/agabrown/astrometry-inference-tutorials}}.
This results in an independent estimate of the distance to each
cluster based on the individual parallaxes and their uncertainties.
The per-cluster final distances are estimated along with their standard
deviations, and shown in Table 1.

The plot in Fig. \ref{Figure3} is the result of the Bayesian analysis performed on the
[DBS2003]5 cluster. This analysis appears to point to a mean concentration around 4 $\pm$ 0.4 kpc, which is 
 where the supposed cluster would be located. 
 
 \begin{table}[!h]
\centering
\caption{Gaia DR2 results.}
\begin{tabular}{l|ccc}
\hline\hline\noalign{\smallskip}
Name & D   & $\mu_{\alpha}$& $\mu_{\delta}$\\
     & [kpc] &  [mas/yr]&  [mas/yr]\\
\hline\noalign{\smallskip}
$[DBS2003]5$ & 4.1$\pm$0.4 & -1.696 & 2.098 \\
$[DBS2003]60$ & 5.7$\pm$0.4 & -2.363 & 1.697 \\
$[DBS2003]98$ & 3.1$\pm$0.3 & -0.983 & -2.921 \\
$[DBS2003]116$ & 9.3$\pm$1.0 & -1.562 & -4.352 \\
$[DBS2003]117 $& 1.6$\pm$0.2 & -0.324 & -1.096 \\
\hline
\end{tabular} 
\label{table1}
\end{table}

\begin{table}[!h]
\centering
\caption{2MASS results.}
\begin{tabular}{l|ccc}
\hline\hline\noalign{\smallskip}
Name & D   & $\mu_{\alpha}$& $\mu_{\delta}$\\
     &  [kpc] & [mas/yr]&  [mas/yr]\\
\hline\noalign{\smallskip}
$[DBS2003]5$ & 4.0$\pm$0.5 & -1.751 & 2.404\\
$[DBS2003]60$ & 5.6$\pm$0.5 & -2.357 & 1.681\\
$[DBS2003]98$ & 2.7$\pm$0.2 & -1.719 & -3.666\\
$[DBS2003]116$ & 10.6$\pm$1.0 & -1.493 & -4.322\\
$[DBS2003]117$ & 1.8$\pm$0.3 & -0.121 & -0.829\\
\hline
\end{tabular}
\label{table2}
\end{table}

\section{Results}
The analysis methodology described here was applied to the five embedded clusters indicated above. For the sake of 
saving space the results for all of them are summarized in Table \ref{table1} for Gaia DR2 data and in Table \ref{table2} for 
the 2MASS JHK filters. The results are slightly different for the distances and proper motions of the clusters, but within 
the associated uncertainties as indicated in the tables. To validate our analysis and by way of example, [DBS2003]5 is also 
recognized as the S305 HII region for which \cite{2014AJ....147...53S}  reported a distance d =~ 5.2$\pm$ 1.4 kpc not far from 
the values shown in the tables. Anyway, the results do not demonstrate the existence of real clusters but rather the presence 
of small number of hot stars in very early stages of evolution at a similar distance exciting the surrounding material. Given this 
fact, ASteCA has not found any synthetic cluster able to mimic the star distribution seen in every case.

\vskip1cm

\begin{acknowledgement}
Based on data acquired at Complejo Astronómico El Leoncito, operated under agreement between the Consejo Nacional 
de Investigaciones Científicas y Técnicas de la República Argentina and the National Universities of La Plata, Córdoba 
and San Juan.
\end{acknowledgement}


\bibliographystyle{baaa}
\small
\bibliography{bibliografia}
 
\end{document}